\documentclass[aps,prb,preprint,showpacs,preprintnumbers,amsmath,amssymb,endfloats]{revtex4}

\usepackage{graphicx,bm,hyperref}

\begin{document}

\title{Exchange energy and stability diagram of few-electron coupled elongated quantum dots}

\author{L.-X. Zhang, D. V. Melnikov, and J.-P. Leburton}
\affiliation{
Beckman Institute for Advanced Science \& Technology and Department of Electrical and Computer Engineering,
University of Illinois at Urbana-Champaign, Urbana, Illinois 61801}

\date{\today}
\begin{flushleft}

\end{flushleft}
\begin{abstract}

We study the properties of a few-electron system confined in 
coupled elongated quantum dots (QDs) 
using a model Gaussian potential and the numerical exact 
diagonalization technique. In the absence of magnetic fields, 
as the aspect ratio $r$ between the QD extensions in the direction 
perpendicular and parallel to the coupling directions increases, 
the exchange energy exhibits a sharp variation at the specific value $r=3.9$, 
before (after) which the exchange
energy increases (declines). The sharp variation occurs
because of a sudden change in the single-particle configuration
of the triplet state. The stability region with one electron in 
each of the QDs is found to shrink, and finally vanishes
as it becomes progressively easier to localize both
electrons into the lower QD. 
For $r>3.9$,
the first singlet-triplet transition shifts to a
small magnetic fields. 
\end{abstract}


\maketitle

\section{Introduction}

Coupled quantum dots (QDs) based on two-dimensional electron 
gas (2DEG) formed with GaAs/AlGaAs heterostructures are promising 
candidates for quantum logic applications because of the 
ability to coherently manipulate the many-body spin states by 
using external electromagnetic fields.\cite{Wiel, Hanson, Loss} 
Recently, a coherent controlled cycle of many-body state preparation, 
spin-interaction and projective read-out has been achieved in 
laterally coupled QDs.\cite{Petta} In such an 
experiment, the electromagnetic control 
of the exchange energy $J$, which drives the Rabi oscillations 
between the 
lowest singlet and triplet states, is of utmost importance. It is well 
known that the 
hyperfine interaction between the electron and nuclear spins
competes 
with the exchange energy to destroy the singlet-triplet 
coherence.\cite{Laird} Therefore, in order to retain the spin-state coherence 
in coupled GaAs/AlGaAs QDs, it is important to optimize the 
exchange energy to exceed the hyperfine interaction significantly.

A wealth of theoretical work has been devoted to study the 
exchange energy in coupled QD 
systems.\cite{Theory_Group,Zhang} The main focus of these studies 
is the tunability of the exchange energy by the electromagnetic 
fields and/or the parameters defining the interdot 
coupling strength, {\it e.g.}, interdot separation and 
barrier height. The optimization of the 
exchange coupling $J$---given a fixed interdot distance, which is
predetermined by the lithography of the top gates---has been rarely discussed. 
In this work, we investigate such a possibility by considering QDs 
elongated perpendicularly to the coupling direction. In this configuration,
one can expect the overlap
between the electron wavefunctions in the two QDs to increase, which
will enhance their interactions.
Our work is encouraged by the recent proposal of
using
coupled elongated QDs to
construct robust spin-qubits with all-electrical qubit
manipulation capabilities.\cite{Kyriakydis}  

In this paper, we perform a detailed
analysis of the 
two-electron system  in coupled elongated QDs 
to show that the exchange coupling indeed becomes larger with
increasing aspect ratio between the extensions of each QD
perpendicular and parallel to the coupling
direction ($r=R_y/R_x$). 
Our analysis based on the numerical exact diagonalization  
technique indicates that the cause of this enhancement is far from
intuitive, while there is an optimum $r$ value beyond which
the exchange energy $J$ decreases. Furthermore, for $r\geq5$, we find 
that the stability region
for one electron in each QD shrink to vanish. Finally, the 
magnetic field, which defines the boundary between
different spin phases of the system ground state, decreases with
increasing $r$.

\section{Model and Method}

The Hamiltonian for the coupled QD system is given by
\begin{equation}
H= H_{orb}+H_Z,
\label{eqn:H_tot}
\end{equation} 
\begin{equation}
H_{orb}=h({\bf r_1})+h({\bf r_2})+C({\bf r_1},{\bf r_2}),
\label{eqn:H_orb}
\end{equation} 
\begin{equation}
h({\bf r}) = \frac{1}{2m^*}({\bf p}+\frac{e}{c}{\bf A})^2 + V({\bf r}),
\label{eqn:H_single}
\end{equation} 
\begin{equation}
C({\bf r_1},{\bf r_2})=e^2/\epsilon|{\bf r_1}-{\bf r_2}|,
\label{eqn:C}
\end{equation} 
\begin{equation}
H_Z = g\mu_B\sum_{i}{\bf B}\cdot{\bf S_i}.
\label{eqn:H_Z}
\end{equation}
\noindent Here, we use the material parameters of GaAs, electron effective 
mass $m^*=0.067m_e$, dielectric constant 
$\epsilon=12.9$, and g-factor $g=-0.44$. $\mu_B$ is 
the Bohr magneton, and ${\bf A} = \frac{1}{2}[-By,Bx,0]$ is the vector 
potential for the constant magnetic field $B$ oriented perpendicular 
to the QD plane ($xy$-plane). The Zeeman effect simply induces
a lowering of the single-particle (SP) and triplet energies by
$13$ and $25$ $\mu$eV/T, respectively. 

We use the following model potential for the coupled QD system:\cite{Zhang_IEEE}
\begin{eqnarray}
V({\bf r})&=&-V_Le^{-(x+d/2)^2/R_x^2+y^2/R_y^2}\nonumber\\
&&-V_Re^{-(x-d/2)^2/R_x^2+y^2/R_y^2},
\label{eqn:potential}
\end{eqnarray}
where $V_L$ and $V_R$ are the depth of the left and right QDs 
(equivalent to the QD gate voltages in experimental structures \cite{Wiel}) 
which can be independently varied, $d$ is the interdot separation, $R_x$ 
and $R_y$ are the radius of the each QD in the $x$ and $y$ direction, respectively.
In this work, we fix $R_x=30$ nm, and define QD {\it aspect
ratio} $r=R_y/R_x$. Numerical exact diagonalization technique is used to solve for the single- and
two-electron energies. Details
of the method are published elsewhere.\cite{Dmm1, Zhang_IEEE} 

Upon completion of the diagonalization procedure, we extract
the SP energies $e_i$ and the two-particle energies $E^{S/T}_i$. 
Here, ``$S$'' (``$T$'') denotes the singlet (triplet) state (In this
paper, if not otherwise mentioned, ``singlet'' and ``triplet''
refer to the singlet and triplet states lowest in energy, respectively). 
The chemical potential of the $N$-th
electron is given by the following equation:~\cite{Wiel}
\begin{equation}
\mu(N) = E_0(N) - E_0(N-1),
\label{eqn:mu}
\end{equation}
where $E_0(N)$ [note $E_0(0)=0$] refers to the ground state 
energy with $N$ electrons in the system. The exchange energy is given by
\begin{equation}
J = E_0^T(2) - E_0^S(2).
\label{eqn:J}
\end{equation}
\noindent For further analysis, the total
energy of the two-electron system is partitioned
into the expectation values of the 
SP energy $K$ and Coulomb energy $C$
\begin{eqnarray} 
E^{S/T}&=&\left\langle \Psi_0^{S/T}|H|\Psi_0^{S/T}\right\rangle\nonumber\\
&=&\left\langle \Psi_0^{S/T}|h({\bf r_1})+h({\bf r_2})|\Psi_0^{S/T}\right\rangle\nonumber\\
&&+\left\langle \Psi_0^{S/T}|C({\bf r_1},{\bf r_2})|\Psi_0^{S/T}\right\rangle\nonumber\\
&=&K^{S/T} + C^{S/T},
\label{eqn:partition}
\end{eqnarray}
\noindent while the spectral function is defined as the projection 
coefficients of the lowest singlet and triplet states  onto 
the SP product states\cite{Dmm2}
\begin{equation} 
\alpha^{S/T}_{k,l} = \left\langle \psi_k({\bf r_1})\psi_l({\bf r_2})|\Psi_0^{S/T}({\bf r_1},{\bf r_2})\right\rangle.
\label{eqn:proj}
\end{equation}
\noindent The electron density is given by
\begin{equation} 
\rho^{S/T}({\bf r_1}) = \int|\Psi_0^{S/T}({\bf r_1},{\bf r_2})|^2d{\bf r_2}. 
\label{c2:eqn:density}
\end{equation}
\noindent Finally, the expectation 
value of the parity operator is given by
\begin{eqnarray}  
\left\langle \hat{P}^{S/T} \right\rangle&=&\left\langle \Psi_0^{S/T}(x_1,y_1,x_2,y_2)\right.\nonumber\\
&&\left| \Psi_0^{S/T}(-x_1,-y_1,-x_2,-y_2) \right\rangle, 
\end{eqnarray}  
and for the parity operator along the  
$y$-axis 
\begin{eqnarray}  
\left\langle \hat{P}^{S/T}_y \right\rangle&=&\left\langle \Psi_0^{S/T}(x_1,y_1,x_2,y_2)\right.\nonumber\\
&&\left| \Psi_0^{S/T}(x_1,-y_1,x_2,-y_2) \right\rangle. 
\end{eqnarray}  

\section{Results}
\subsection{Aspect ratio dependence of the 
exchange energy}

\begin{figure}[tb]
\includegraphics[width=7cm]{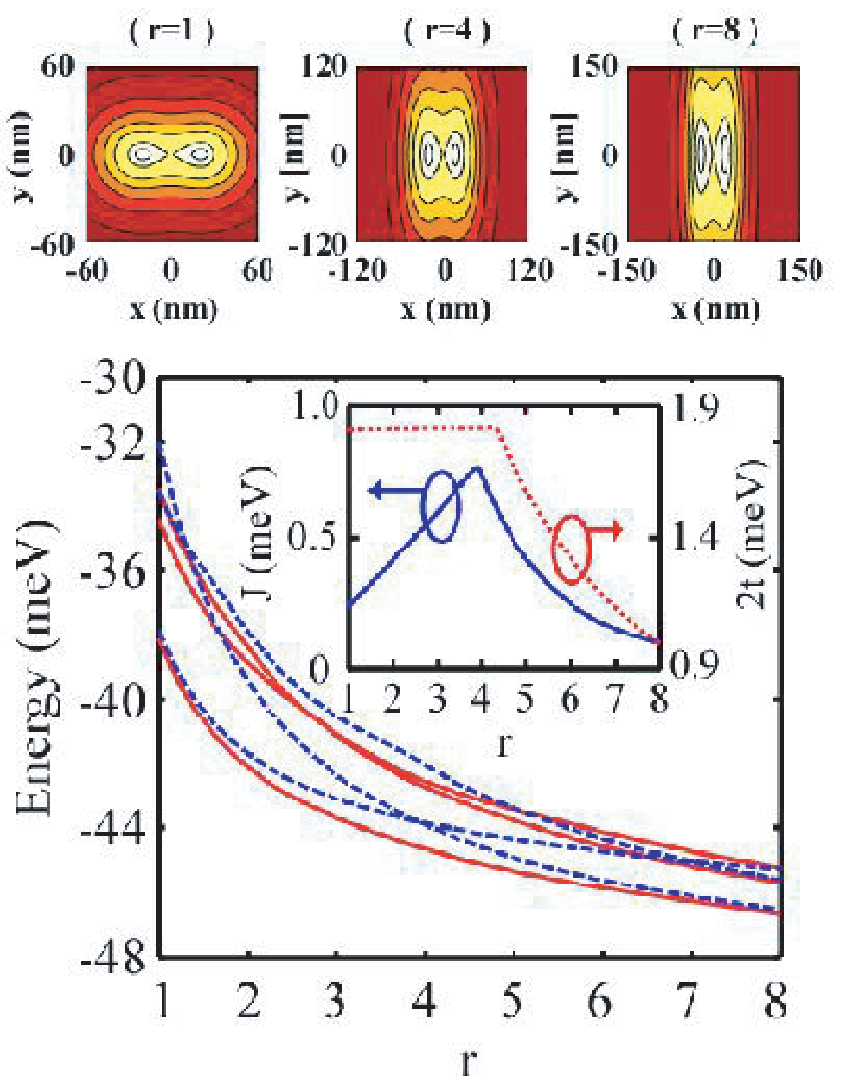}
\caption{\label{fig:fig1} (Color online) Top panels: potential contour 
plots of coupled QD with $r=1$ (left), $r=4$ (middle), 
and $r=8$ (right). Redder (darker gray) regions 
correspond to higher potential. Bottom panel: three 
lowest singlet (red/gray, solid lines) and triplet (blue/dark gray, 
dashed lines) energy levels as a function of QD aspect 
ratio $r$. The inset shows 
$r$ dependence of the exchange energy $J$ 
(blue/dark gray, solid) and tunnel coupling $2t$ (red/gray, dotted). For all panels,
$V_L=V_R=25$ meV, $d=50$ nm, $B=0$ T.}
\end{figure}

Figure~\ref{fig:fig1} top panels show the potential 
contour plots $r=1$ (left), $r=4$ (middle), and $r=8$ (right). As $r$ 
increases, the potential becomes more elongated in the 
$y$-direction, while the effective interdot distance ({\it i.e.}, 
the $x$-distance between the two minima of the potential) 
and the interdot barrier height remain 
constant at $40$ nm and $1.98$ meV, respectively.

In the lower panel of Fig.~\ref{fig:fig1}, we plot the three lowest 
singlet (red/gray, solid) and triplet (blue/dark gray, dashed) energy levels 
as a function of $r$. With increasing $r$, the SP 
energies decreases (not shown), resulting in the decrease of 
the two-particle energy levels. We note that the lowest energy 
of the singlet state [$E_0^S(2)$] decreases smoothly with $r$, 
while the lowest energy of the triplet state [$E_0^T(2)$] exhibits 
a cusp at $r=3.9$ because of the crossing of the lowest two 
triplet state energy levels.  This cusp results in a sharp
variation in the exchange energy dependence on $r$, which
is shown in the inset of the lower panel of Fig. \ref{fig:fig1}. 
In the same inset,
we show the variation of the tunnel coupling $2t=e_1-e_0$.  
For $r\leq4.3$, 
the SP ground and 
first excited states have $s$ and $p_x$ characters, respectively, 
and $2t$ barely increases from $1.8105$ to $1.8114$ meV with 
increasing $r$, because the energy contributions from the $y$-direction 
to $e_0$ and $e_1$ cancel out. For $r>4.3$, the SP 
first excited state bears a $p_y$ character, which causes $2t$ to 
decrease monotonically with $r$. 

\begin{figure}[tb]
\begin{center}
\includegraphics[width=7cm]{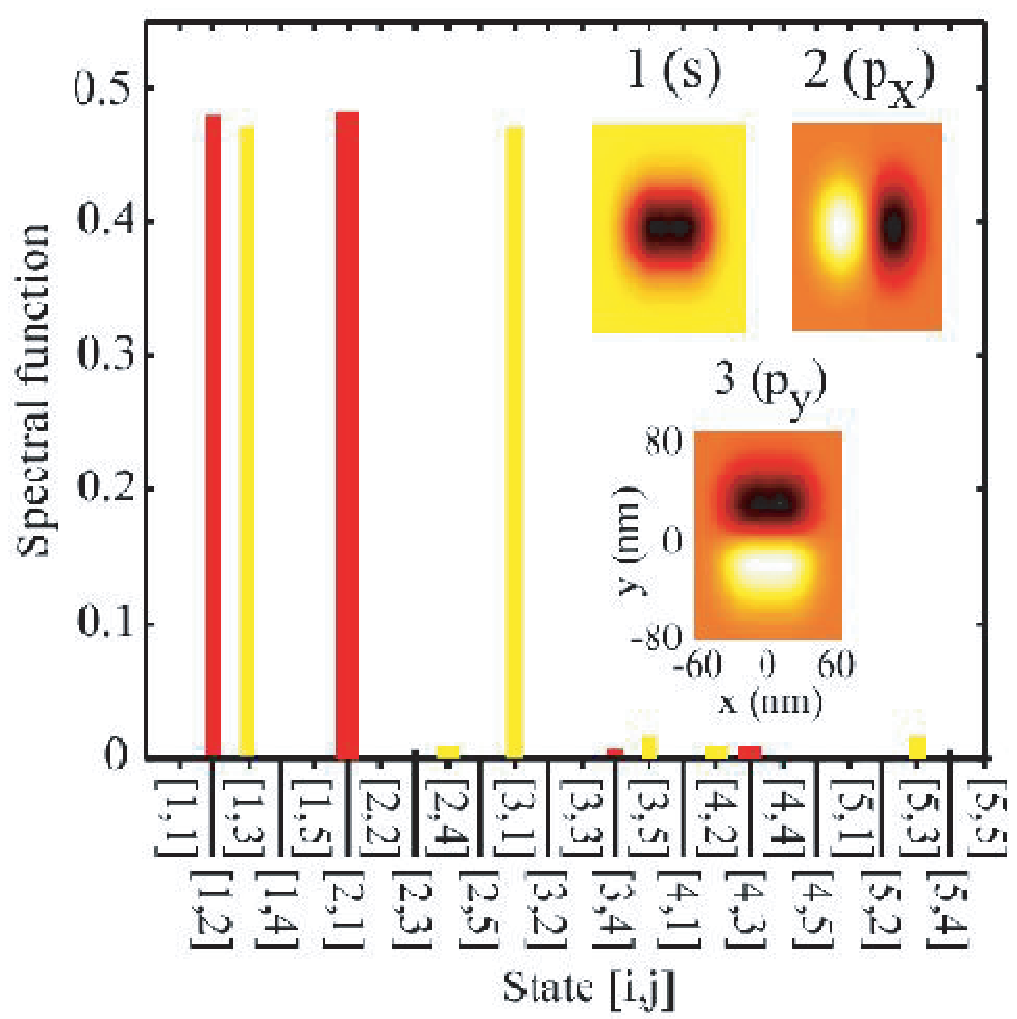}
\caption{(Color online) Spectral decomposition of the two-electron
wavefunction onto different single-particle pairs.
The red (dark) columns are for $r=3.9$, while
the yellow (bright) columns are for $r=4$. The inset
shows the contour plots of lowest three 
single-particle states in ascending 
order (indicated by number) of energy for 
both $r=3.9$ and $r=4$. The state symmetry 
is shown in parenthesis.}
\label{fig:fig2} 
\end{center}
\end{figure}

In order to investigate in detail the cusp
in the lowest triplet state energy, or, the crossing
between the two lowest triple levels in the lower
panel of Fig.~\ref{fig:fig1}, 
we plot in Fig.~\ref{fig:fig2} the spectral function
of the two-electron wavefunction.
It is seen that at $r=3.9$
the triplet mainly consists of the $[1,2]$ and $[2,1]$ 
SP state pair, while at $r=4$ it mainly consists of
the $[1,3]$  and $[3,1]$ SP state pair. Here, $1$, $2$ and $3$
denote the SP states in ascending energy, which have
$s$, $p_x$ and $p_y$ characters, respectively, as shown
in the Fig.~\ref{fig:fig2} inset. Since the
energy ordering of these SP states 
does not change as $r$ changes from $3.9$
to $4$ (not shown here), the cusp in the
lowest triplet state is due to a sudden transition of
the triplet wavefunction from occupying an $sp_x$ pair
to an $sp_y$ pair.

\begin{figure}[tb]
\begin{center}
\includegraphics[width=7cm]{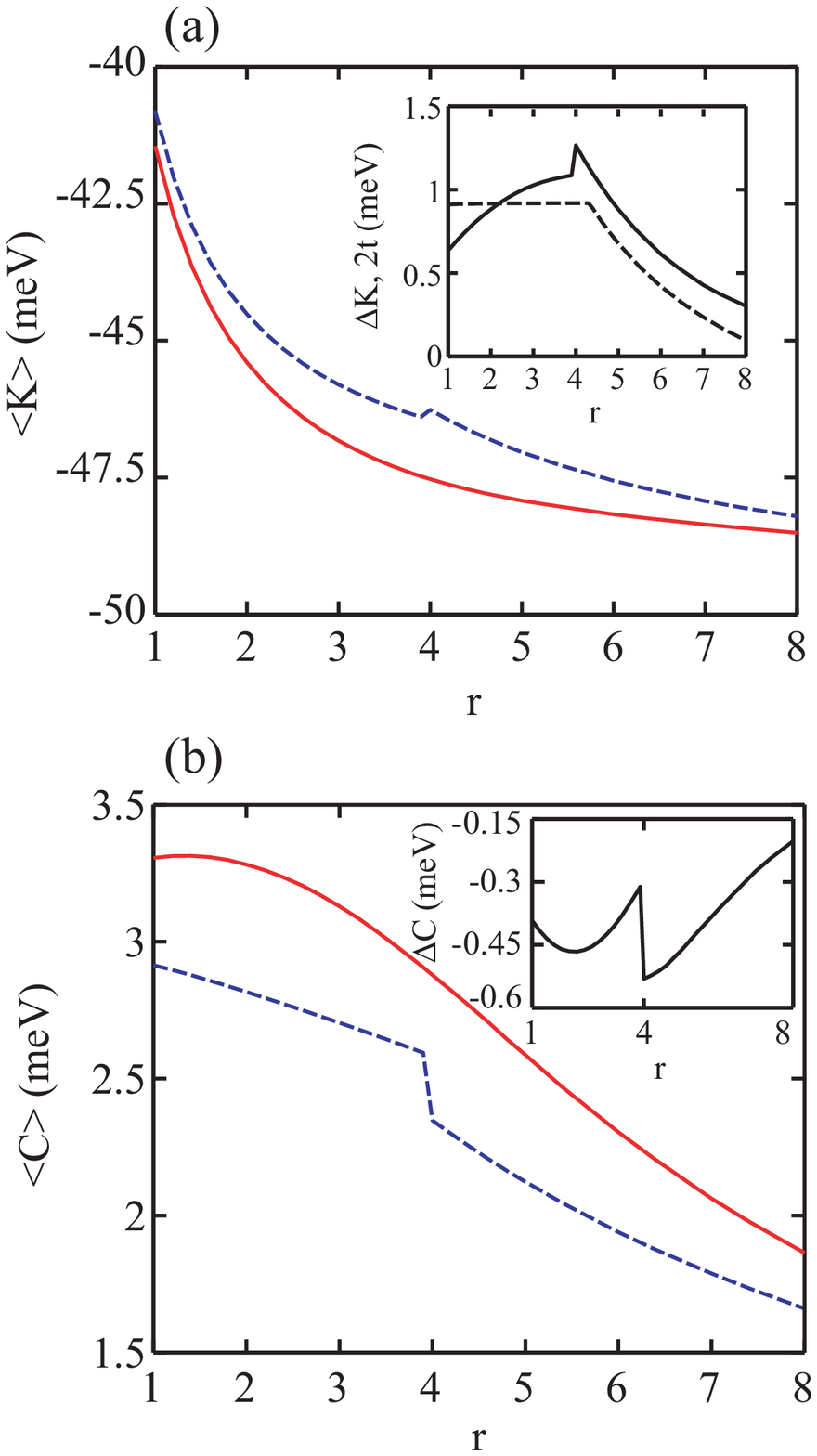}
\caption{(Color online) (a) Single-particle energy contribution
$\left\langle K \right\rangle$ as a function
of QD aspect ratio $r$. The red/gray, solid (blue/dark gray, dashed) line
is for the singlet (triplet) state. 
Inset: the solid line shows
the difference 
$\Delta K = \left\langle K^T \right\rangle -\left\langle K^S \right\rangle $.
The dashed line shows $2t$ as a comparison.
(b) Coulomb energy contribution
$\left\langle C \right\rangle$ as a function
of QD aspect ratio $r$.  The red/gray, solid (blue/dark gray, dashed) line
is for the singlet (triplet) state. 
Inset: the solid line shows
the difference
$\Delta C = \left\langle C^T \right\rangle -\left\langle C^S \right\rangle $.
\label{fig:fig3}} 
\end{center}
\end{figure}

In Fig.~\ref{fig:fig3}, we plot separately the
(a) SP $<K>$ and (b) Coulomb $<C>$ contributions to 
the singlet and triplet state energies as a function
of the QD aspect ratio $r$. As $r$ increases, the general trend
for all these energy terms is to decrease, leading to 
decreasing singlet and triplet energies shown in Fig.~\ref{fig:fig1}.
For the singlet state, both $<K>$ and $<C>$ terms decrease 
smoothly with $r$. For the triplet state, however,
a discontinuity is seen from $r=3.9$ to $4$: $<K>$ ($<C>$) suddenly
increases (decreases) by $0.128$ ($0.607$) meV. It now
becomes clear that the transition of the SP configuration 
shown in Fig.~\ref{fig:fig2} from the $sp_x$ pair 
to the $sp_y$ pair is favored by the lowering of the Coulomb
interaction despite the increase in the SP energy.
The insets in Fig.~\ref{fig:fig3}(a) and (b) show that
the difference in the Coulomb energy between singlet and triplet
states ($\Delta C = \left\langle C^T \right\rangle -\left\langle C^S \right\rangle <0$) 
is always overcome by the SP 
energy contribution  
($\Delta K = \left\langle K^T \right\rangle -\left\langle K^S \right\rangle >0$), 
leading to a positive exchange interaction
($J=\Delta K +\Delta C $, see Fig.~\ref{fig:fig1}).\cite{footnote1} 
The comparison between $\Delta K$ and the
tunnel coupling $<2t>$ in the inset of Fig.~\ref{fig:fig3}(a) 
shows that the SP energy contribution to the
singlet and triplet states is strongly influenced by
the Coulomb interaction and is quite different from
the noninteracting picture.  

As a consequence of the sudden change in the 
SP occupation, the $y$-symmetry $P_y$ of 
the two-electron wavefunction of the lowest
triplet state changes abruptly from $1$ to $-1$, which
is validated by direct calculation of $P_y$. We point
out that the crossing between the lowest two
triplet states by increasing $r$ is allowed 
because they possess opposite
$y$-symmetry, which exemplifies the general
von Neumann-Wigner theorem relating the molecular
energy levels to the two-electron wavefunction 
symmetry.\cite{Wigner}

\begin{figure}[t]
\begin{center}
\includegraphics[width=7cm]{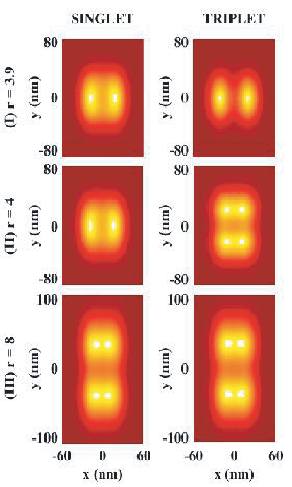}
\caption{(Color online) Contour plots of the electron density
for both singlet (left column) and triplet (right column).
Rows I, II and III are for $r=3.9$, $r=4$ and $r=8$,
respectively. In the plots, redder (darker gray) regions correspond
to lower electron density.}
\label{fig:fig4} 
\end{center}
\end{figure}

\begin{figure*}[tb]
\includegraphics[width=14.5cm]{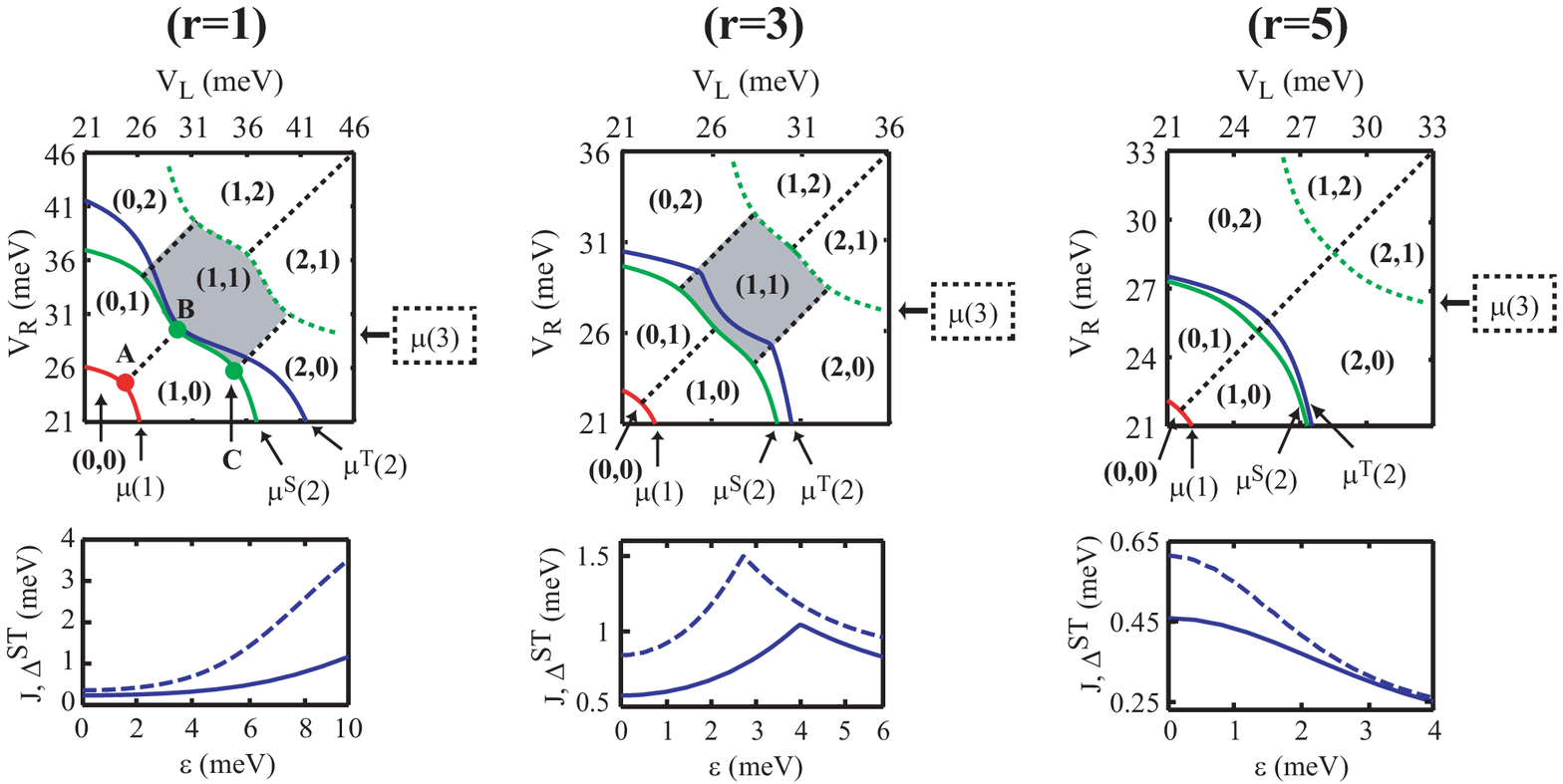}
\caption{\label{fig:fig5}(Color online) Top panels: stability 
diagrams for $r=1$ (left), $r=3$ (middle) 
and $r=5$ (right). In each diagram, the red, green 
and blue curves (solid) are computed contour lines 
at which the chemical potential $\mu(1)$, $\mu^S(2)$ 
and $\mu^T(2)$ equal to the reference 
value $\mu_{ref}=-21$ meV, respectively. Curves 
for different chemical potentials are also 
indicated by arrows. The dotted straight lines are a 
guide for eyes separating different stable charge 
states. Note that the exact locations of 
the $\mu(3)$ curve (green dotted curve) 
and $(1,2)$, $(2,1)$ regions 
are not computed. In the left two top panels, 
the $(1,1)$ region is indicated by the shaded area.
In the left most upper panel, 
we also indicate the double-triple points A and B. 
Point C is where the $\mu^S(2)$ curve has the largest 
curvature for $V_L \neq V_R$. For corresponding QD aspect ratios, 
the bottom panels
show $J$ (solid curves) 
as a function of interdot detuning $\epsilon=V_L-V_R$
from the center
of the $(1,1)$ region.
The dashed curves on the bottom panels show the 
separation ($\Delta^{ST}$) between the contour 
lines of $\mu^S(2)$ and $\mu^T(2)$ 
projected along the main diagonal of as a function of interdot 
detuning $\epsilon=V_L-V_R$. All data are obtained
at $R_x=30$ nm, $d=50$ nm and $B=0$ T.}
\end{figure*}

The contour plots in Fig.~\ref{fig:fig4} clearly show
that from $r=3.9$ (first row) to $r=4$ (second row)
the electron density in the lowest singlet state barely
changes, while the density in the lowest triplet
state changes abruptly from two peaks localized in the 
left and right QDs (the separation of two peaks in 
the $x$-direction is $\sim40$ nm) to four peaks 
separated along both $x$ and $y$ 
directions (separation between
peaks in the $x$ and $y$ directions are
$20$ and $40$ nm, respectively), 
again due to the sudden change in
the SP configuration. 
The third row in Fig.~\ref{fig:fig3} shows that at 
$r=8$, both the singlet and triplet densities 
exhibit four peaks separated in both the $x$ 
and $y$ directions. Our analysis shows that 
from $r=4$ to $r=8$, the left and right peaks 
in the singlet state density gradually separate 
into four peaks, and the separation between 
the top two and bottom two peaks in the triplet 
state density smoothly increases. Such electron
localization effects at large $r$ are discussed
for other many-electron 
QD systems with weak confinement, see, 
{\it e.g.} Ref.~\onlinecite{Bednarek} and
references therein. 

\subsection{Stability diagrams}

In Fig. \ref{fig:fig5}, upper panels, we plot the 
stability diagrams \cite{Wiel} of the coupled 
QDs for $r=1$ (left), $r=3$ (middle), and $r=5$ 
(right) for $R_x=30$ nm, $d=50$ nm and $B=0$ T. 
The solid curves indicated by arrows shows 
the computed contours, where chemical potentials of 
the first electron (red), the second electron in the 
singlet state (green), and second electron in the 
triplet state (blue) are equal to the reference 
value [$\mu(1)=\mu^S(2)=\mu^T(2)=-21$ meV]. According 
to the general shape of the stability diagram for 
coupled QDs,\cite{Wiel} we use dotted straight 
lines on the diagrams to separate different charge 
states indicated by discrete electron numbers
on the left and right QDs, 
{\it e.g.}, $(0,1)$ means zero electrons on 
the left QD and one electron on the right QD. 
Specifically, the boundaries between the $(1,1)$ 
and $(0,2)$ [or $(2,0)$] states are taken extending 
from the point on the $\mu^S(2)$ curve at which the 
curvature is the largest for $V_L \neq V_R$, {\it e.g.}, 
point C on the upper left panel and parallel to the 
main diagonal. In the absence of magnetic field ($B=0$), 
the $\mu^S(2)$ curve is the boundary between one and 
two electrons in the system (in the linear transport 
regime wherein the source and drain chemical potentials 
are nearly the same).\cite{footnote1} Based on this fact, 
we extrapolate from the first off-diagonal triple 
point ({\it e.g.}, point C on the upper left panel) to 
get the boundary between two- and three-electron 
states [green dotted curve indicated by $\mu(3)$].
Here, we assume that the triple point separation 
between charge states $(1,0)$ and $(2,1)$ 
(or between the $(0,1)$ and $(1,2)$) is the same
as the separation between the $(0,0)$ and $(1,1)$ states.\cite{footnote2}

\begin{figure*}[tb]
\includegraphics[width=14.5cm]{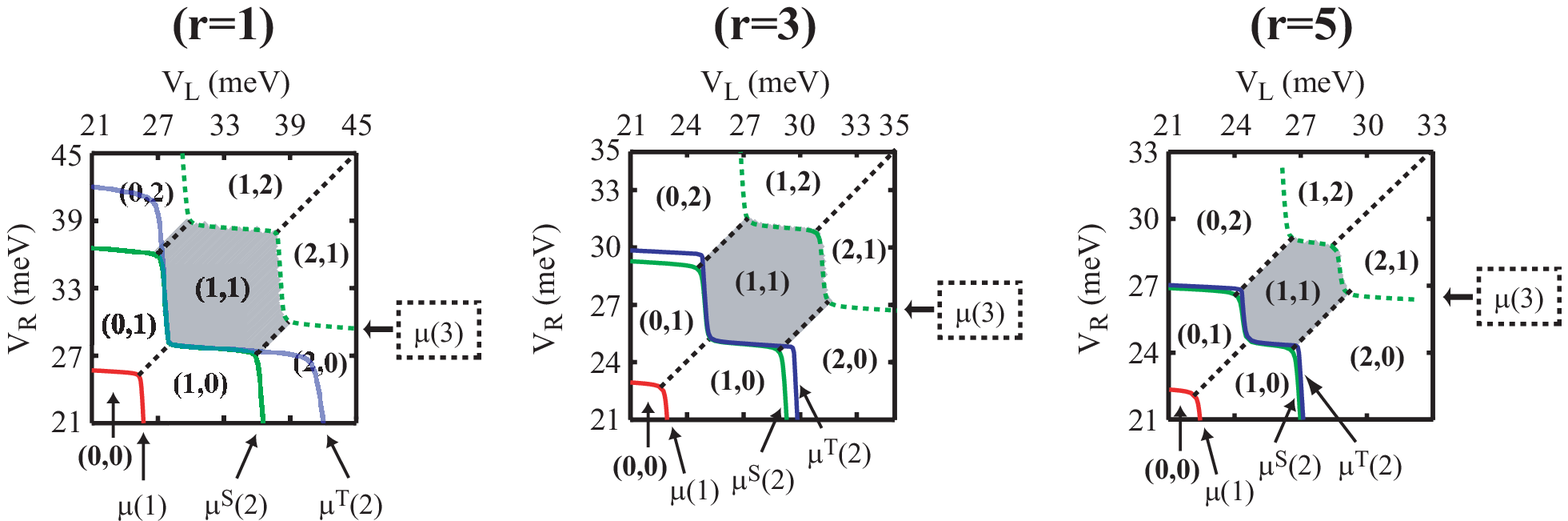}
\caption{\label{fig:fig6}(Color online) Stability 
diagrams for $r=1$ (left), $r=3$ (middle) 
and $r=5$ (right). In each diagram, the red, green 
and blue curves (solid) are computed contour lines 
at which the chemical potential $\mu(1)$, $\mu^S(2)$ 
and $\mu^T(2)$ equal to the reference 
value $\mu_{ref}=-19$ meV, respectively. Curves 
for different chemical potentials are also 
indicated by arrows. The dotted straight lines are a 
guide for eyes separating different stable charge 
states. Note that the exact locations of 
the $\mu(3)$ curve (green dotted curve) 
and $(1,2)$, $(2,1)$ regions 
are not computed. In each panel, 
the $(1,1)$ region is indicated by the shaded area. 
All data are obtained
at $R_x=30$ nm, $d=60$ nm, and $B=0$ T.}
\end{figure*}

In Fig. \ref{fig:fig5}, we notice that, on the one hand, as
$r$ increases, the crossing points of the 
$\mu(1)$, $\mu^S(2)$ and $\mu^T(2)$ curves with the 
main diagonal shift to smaller $V_L=V_R$ values 
because the SP energies decreases as $r$ increases,
and as such a less negative $V_L$ and $V_R$ value is 
required to charge the coupled QDs.  
On the other hand, the double-triple point (DTP) 
separation, {\it i.e.}, the separation between the 
crossing points of $\mu(1)$ and $\mu(2)$ curves with 
$V_L=V_R$, decreases with $r$. For the singlet (triplet) 
state, the DTP separation measured in $\Delta V_L=\Delta V_R $ 
is $5.181$ ($5.269$), $4.128$ ($4.725$) and $3.473$ ($3.907$) 
for $r=1$, $3$ and $5$, respectively. This decreasing trend
of the DTP separation suggests that the coupling strength
between the two QDs decreases with 
increasing $r$ (see Ref.~\onlinecite{Wiel}). However,
from our direct calculations shown in
the inset of Fig. \ref{fig:fig1}, lower
panel, the tunnel coupling decreases only for $r>4.3$, while the
exchange energy is largest for $r=3.9$. The discrepancies
regarding the coupling strength 
between the DTP separation and direct calculations of the 
tunnel and exchange couplings can be understood by observing the
following: the DTP
separation is given by $2t+C$, where $2t$ and $C$ denote tunnel
coupling and interdot Coulomb interaction. As $r$ increases
$2t$ decreases for $r>4.3$, while $<C>$ monotonically 
decreases for both singlet and triplet (see Fig. \ref{fig:fig3}). 
As a result, the DTP separation decreases. The exchange coupling, however,
is determined by the  energy {\it difference} between the singlet
and triplet states. As shown in Fig. \ref{fig:fig3},
such energy difference, when splitted into the SP
contribution $<\Delta K>$ and the Coulomb contribution $<\Delta C>$,
has a complicated dependence on $r$. 
In contrast, if the interdot separation were increased to decouple the
two QDs, then all quantities $2t$, $C$, $<\Delta K>$, and $<\Delta C>$
would decrease, leading to both decreasing DTP separation and exchange
energy.\cite{Dmm2,Zhang}  
 
One important feature shown in Fig. \ref{fig:fig5} is that as
$r$ increases, the distance between the triple points on the
main diagonal and the first off diagonal ({\it e.g.},
points B and C in the upper left panel of Fig. \ref{fig:fig5}) 
becomes
smaller, and at large $r$ these triple points coincide. Consequently,
the $(1,1)$ stability region shrinks and finally disappears. This
is because at large aspect ratios, even a small amount of
interdot detuning can localized both electrons into
the lower QD, resulting in an unstable $(1,1)$ charge state. 
The boundary $\mu^T(2)$ at $r=5$ suggests that the
$(1,1)$ charge state is also unstable for the triplet state, although
the $\mu^S(2)$ and $\mu^T(2)$ curves evolve in different
fashion as $r$ increases.

After locating the different charge stable regions 
on the stability diagram, we now investigate 
the interdot detuning effect by departing from 
the center of the $(1,1)$ region along the direction 
perpendicular to the main diagonal, {\it i.e.},
$V_L+V_R=constant$. Such detuning effects are 
important as two electrons transfer to a single 
QD, which is a key step in spin coherent 
manipulation and spin-to-charge conversion 
in two-electron double QD experiments for quantum 
logic gate applications.\cite{Hanson, Petta} 

The solid curves in Fig.~\ref{fig:fig5}, 
lower panels, show
the exchange energy $J$ as a function of interdot
detuning $\epsilon=V_L-V_R$ along the $V_L+V_R=constant$
line [$\epsilon=0$ is chosen at the $(1,1)$ region
center]. In the case of coupled circular
QDs ($r=1$), both singlet
and triplet states localize progressively into the
lower QD with increasing $\epsilon$, leading to a monotonic increase of $J$.
Such a dependence is similar to recent 
experimental\cite{Petta} and theoretical\cite{Stopa} results.
For $r=3$, a sharp cusp in $J$ occurs at $\epsilon\sim4$ meV 
before which $J$ monotonically increases with $\epsilon$.
This cusp is induced by a sudden SP
configuration change in the lowest triplet state, which
is similar to the effects seen in Fig.~\ref{fig:fig1} 
and analyzed in Fig.~\ref{fig:fig2}, albeit here the
perturbation in the Hamiltonian is introduced by
interdot detuning instead of deformation effects.
More detailed analysis of the two-particle
energies and electron density for the $r=1$ and $r=3$
cases can be found in Ref.~\onlinecite{Zhang1}.
For $r=5$, we observe that the exchange energy decreases
monotonically with $\epsilon$, because 
the Coulomb energy difference between the singlet
and triplet states becomes smaller as the two
electrons in both the singlet and triplet states localize
at the opposite ends of the lower single QD to minimize
their Coulomb interaction.

In the lower panels of Fig.~\ref{fig:fig5}, we also plot 
the $\epsilon$ dependence of $\Delta^{ST}$ (dashed curves),
the difference between the $\mu^S(2)$ and  $\mu^T(2)$ curves
projected along the main diagonal. $\Delta^{ST}$ is relevant
in this context because in coupled QD experiments the
chemical potential contour lines are mapped out by single-electron
charging measurements, which provides useful information on 
the electronic structure of the QD.\cite{Wiel, Hanson, Johnson} 
Here, we notice that although the general detuning 
dependence is similar between $J$ and $\Delta^{ST}$, a
linear factor is not sufficient to scale values of $J$ 
to overlap with those of $\Delta^{ST}$ because
the two quantities are extracted under different
bias conditions. It should be pointed out that transport experiments
measure the quantity $\Delta^{ST}$, which differs quantitatively 
from the exchange energy $J$.

In Fig. \ref{fig:fig6}, we plot the charge stability
diagram of the coupled QDs for $r=1$ (left), 
$r=3$ (middle), and $r=5$ (right) for 
$R_x=30$ nm, $d=60$ nm and $B=0$ T. Compared to
the data in Fig. \ref{fig:fig5}, which correspond
to strongly coupled QDs,  the data in 
Fig.~\ref{fig:fig6} depict the situation in decoupled
QDs.\cite{Zhang} In this case,
as $r$ increases, (1) the crossing points of the 
$\mu(1)$, $\mu^S(2)$ and $\mu^T(2)$ curves with the 
main diagonal shift to smaller $V_L=V_R$ values; (2)
the DTP separation decreases [The DTP separation
is $2.847$ ($2.860$), $2.618$ ($2.641$), $2.514$ ($2.545$),
for $r=1$, $r=3$, and $r=5$, respectively]; 
and (3) the $(1,1)$ region becomes smaller. These behaviors
are similar to those for $d=50$ nm. However, the $(1,1)$
region does not vanish at $d=60$ nm and $r=5$ because
as the QDs are more decoupled, both the interdot distance
and interdot barrier height become larger, which 
require a larger interdot detuning to ``push'' both electrons 
into the lower QD. At a fixed $r$, the DTP separation
(curvature at the triplet points) is smaller (larger)
for $d=60$ nm than for $d=50$ nm, indicating that
both tunnel coupling and Coulomb interaction are smaller
for more decoupled QDs.\cite{Zhang}

\subsection{Spin phase diagram}

\begin{figure}[tb]
\begin{center}
\includegraphics[width=7cm]{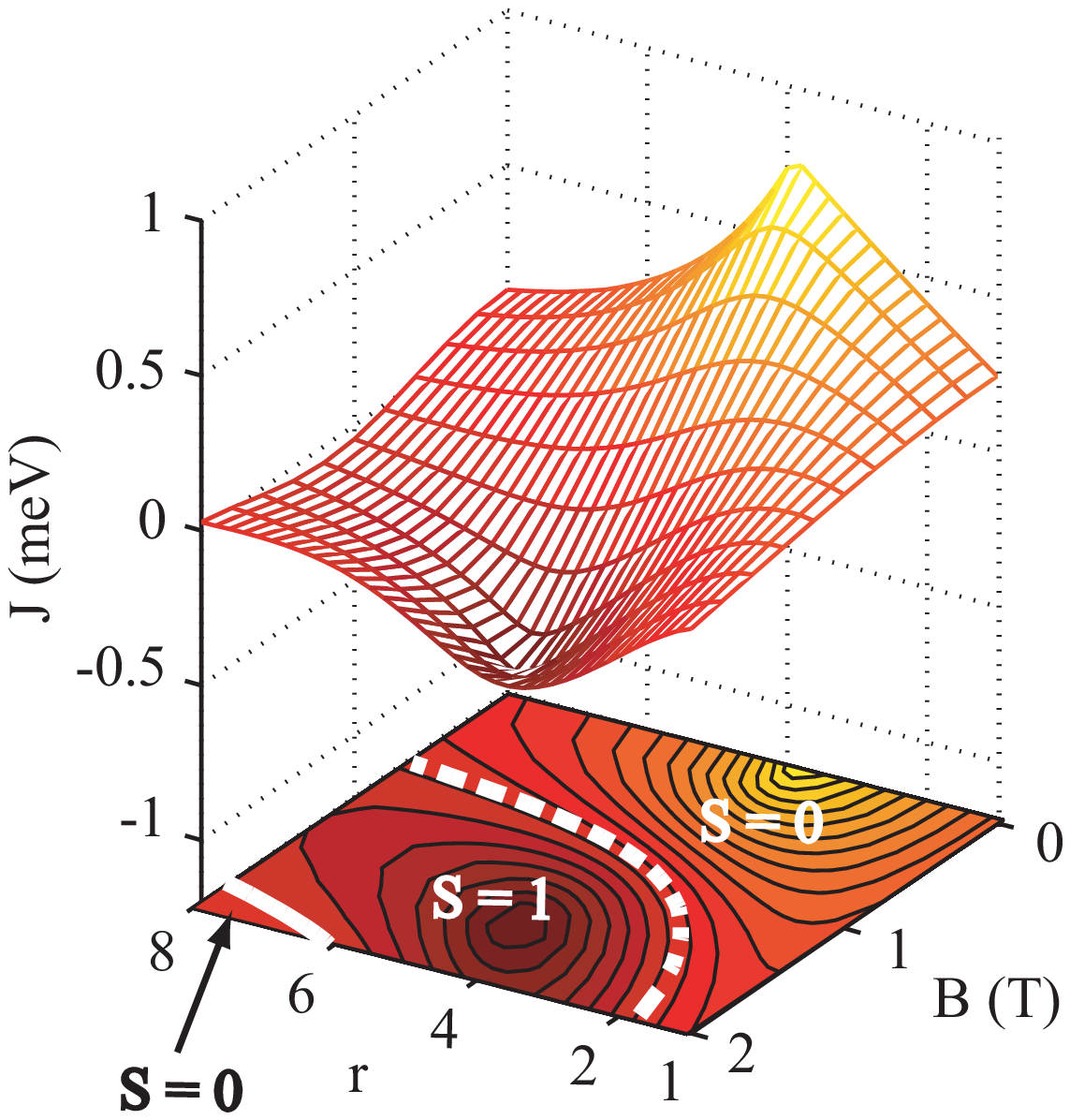}
\caption{\label{fig:fig7}(Color online) Mesh (contour) plot 
of the exchange energy $J$ as a function of QD 
aspect ratio $r$ and the magnetic field $B$. 
The thick white curves (solid and dashed) on 
the contour plot correspond to $J=0$. Total
spin of the two-electron ground state is given in
different regions. Redder (darker gray) regions
correspond to lower $J$ value.}
\end{center}
\end{figure}

In this subsection, we discuss the variation of
the exchange energy $J$ as
a function of both $r$ and $B$. By identifying the
regions where $J$ assumes different signs,
we construct the spin phase diagram in which 
the two-electron ground state spin state
(either $S=0$ or $S=1$) is shown as a 
function of $r$ and $B$.\cite{Harju, Helle}

In Fig. \ref{fig:fig7}, we plot the exchange energy $J$ 
as a function of the QD aspect ratio $r$ and the magnetic 
field $B$ perpendicular to the $xy$-plane. 
At fixed $r$, as $B$ increases, $J$ decreases from its
value at $B=0$ T to become negative and saturate
at very large magnetic field, as previously reported.\cite{Theory_Group}
We note that 
at intermediate $r$ ($r\sim4$), $J$ changes much faster with 
$B$ than at small or large $r$. This 
$B$-field effect at intermediate $r$ 
values is associated with the 2D confinement of the 
QDs,  {\it i.e.}, near $r=4$ the SP level separations 
in the $x$- and $y$-directions are comparable (cf. 
Fig. \ref{fig:fig1}, lower inset, $2t$ curve). 
We also note that, with 
increasing $r$, 
the relative change of $J$ is small for $B\sim1$ T, while 
it is much larger for $B\sim0$ T or $B\sim2$ T. 
The kink in $J$ at $B=0$ T (cf. 
Fig. \ref{fig:fig1}, lower inset, $J$ curve), due to the crossing of two 
lowest triplet levels, does not exist for $B~\neq~0$~T
because a nonzero magnetic field couples the SP states
with different Cartesian symmetries, thereby removing
the condition for the crossing of the lowest two triplet
states. In the investigated ranges of $r$ 
and $B$, $J$ assumes a maximum (minimum) value 
of $0.773$ ($-0.372$) meV at $r\approx3.9$, 
$B\approx0$ T ($r\approx4.4$, $B\approx1.6$ T).

The projected contour plots in Fig.~\ref{fig:fig7} shows 
that the first singlet-triplet transition 
(at which $J$ first crosses zero as $B$ 
increases from zero at fixed $r$) occurs
at a smaller $B$ value as $r$ increases,
which is shown by the thick 
white dashed curve on the contour plot in 
Fig. \ref{fig:fig7}. Such a dependence can be 
understood by observing that, in the absence of
the $B$ field, as $r$ increases the SP
energy spacing decreases, and, for a larger
$r$, a smaller
magnetic field is needed to further decrease the
SP spacing and bring the triplet state
to the ground state with the aid of the Coulomb energy 
difference between the singlet and triplet states. 
At higher magnetic field and 
larger $r$, we observe another contour line for $J=0$
(thick solid white curve at the 
lower left corner). The reappearance of the singlet
state as the ground state is reminiscent of the 
singlet-triplet oscillation found for a two-electron
single QD and also reported elsewhere
for two-electron QDs with strong 
confinement.\cite{Harju, Helle, Wagner} In the
foregoing discussion, we had not included the 
Zeeman energy for the triplet state, 
which would lower the triplet 
energy such that the boundary 
for the first singlet-triplet transition (thick white dashed
curve) would shift
to lower values of $r$ and $B$, while
the second singlet-triplet transition (thick white solid
curve) would move to higher values of $r$ and $B$.

\section{Conclusions}

We have shown that the exchange energy between
two electrons in coupled elongated quantum dots
is enhanced by increasing the aspect ratio of the 
dots in the perpendicular direction to the coupling
direction. 
However, there is an optimum aspect ratio
beyond which the electron density in each dot starts
to localize, and the exchange energy decreases.
With increasing aspect ratio, 
the $(1,1)$ region becomes unstable with respect
to interdot detuning, which is undesirable
for two spin-qubit operations. We have also
shown that the exchange energy in symmetrically
biased coupled quantum dots
is tunable between
maximum (positive) and minimum (negative) values
 by varying
the magnetic field and the QD aspect ratio.

\begin{acknowledgments}
This work is supported by the DARPA QUIST program 
and NSF through the Material Computational Center at the 
University of Illinois. LXZ 
thanks the Beckman Institute, Computer Science 
and Engineering program, and the Research Council 
at the University of Illinois.
\end{acknowledgments}

\end{document}